\documentclass{mn2e}
\usepackage{graphicx}
\usepackage{breqn}

\def\mujybeam{$\mu$Jy$\,$beam$^{-1}$\ }

\title[Central images in CLASS~B1030+074]
{A new VLA/e-MERLIN limit on central images in the gravitational lens system CLASS B1030+074}
\author[Quinn et al.]
{Jonathan Quinn$^{1}$, Neal Jackson$^{1}$, Amitpal Tagore$^{1}$, Andrew Biggs$^{2}$, \\
\mbox{}\\
{\rm \LARGE Mark Birkinshaw$^{3}$, Scott Chapman$^{4}$, Gianfranco De Zotti$^{5,6}$, John McKean$^{7,8}$,}\\
\mbox{}\\
{\rm \LARGE Ismael P\'erez-Fournon$^{9,10}$, Douglas Scott$^{11}$, Stephen Serjeant$^{12}$}\\ 
\mbox{}\\
$^{1}$Jodrell Bank Centre for Astrophysics, School of Physics \& Astronomy, 
University of Manchester,  Oxford Road, Manchester M13 9PL\\
$^{2}$ESO, Karl-Schwarzschild-Strasse 2, 85748 Garching bei M\"unchen, Germany\\
$^{3}$University of Bristol, School of Physics, Tyndall Avenue, Bristol BS8 1TL\\
$^{4}$Department of Physics and Atmospheric Science, Dalhousie University, Halifax, NS B3H 3J5, Canada\\
$^{5}$INAF, Osservatorio Astronomico di Padova, Vicolo dell'Osservatorio 5, I-35122 Padova, Italy\\
$^{6}$SISSA, via Bonomea 265, I-34136 Trieste, Italy
$^{7}$ASTRON, Oude Hoogeveensedijk 4, 7990 AA Dwingeloo, Netherlands\\
$^{8}$Kapteyn Institute, University of Groningen, PO Box 800, 9700 AV Groningen, Netherlands\\
$^{9}$Instituto de Astrof\'{\i}sica de Canarias, E-38205 La Laguna, Tenerife, Spain\\
$^{10}$Universidad de La Laguna, Dpto. Astrofísica, E-38206 La Laguna, Tenerife, Spain\\
$^{11}$Department of Physics \& Astronomy, University of British Columbia, 6224 Agricultural Road, Vancouver, BC V6T 1Z1, Canada\\
$^{12}$Department of Physical Sciences, The Open University, Milton Keynes, MK7 6AA\\
}
\begin{document}
\maketitle
\begin{abstract}
We present new VLA 22-GHz and e-MERLIN 5-GHz observations of CLASS~B1030+074, a two-image strong gravitational lens system whose background source is a compact flat-spectrum radio quasar. In such systems we expect a third image of the background source to form close to the centre of the lensing galaxy. The existence and brightness of such images is important for investigation of the central mass distributions of lensing galaxies, but only one secure detection has been made so far in a galaxy-scale lens system. The noise levels achieved in our new B1030+074 images reach 3~\mujybeam and represent an improvement in central image constraints of nearly an order of magnitude over previous work, with correspondingly better resulting limits on the shape of the central mass profile of the lensing galaxy. Simple models with an isothermal outer power law slope now require either the influence of a central supermassive black hole, or an inner power law slope very close to isothermal, in order to suppress the central image below our detection limit. Using the central mass profiles inferred from light distributions in Virgo galaxies, moved to $z=0.5$, and matching to the observed Einstein radius, we now find that 45\% of such mass profiles should give observable central images, 10\% should give central images with a flux density still below our limit, and the remaining systems have extreme demagnification produced by the central SMBH. Further observations of similar objects will therefore allow proper statistical constraints to be placed on the central properties of elliptical galaxies at high redshift.
\end{abstract}


\large

\section{Introduction}

Strong gravitational lens systems, in which a background object is multiply imaged by a foreground mass, carry important information about mass distributions in galaxies at cosmological distances. In most cases, a lens system in which two objects lie close enough along the same line of sight consists of an even number of visible images, corresponding to stationary points on the Fermat surface (e.g. Schneider, Ehlers \& Falco 1992). In a two-image system\footnote{We use ``two-image'' and ``four-image'' to refer to lens systems containing two or four images on kiloparsec scales.}, these consist of a bright image at the Fermat minimum, and a generally weaker second image at a saddle point; in a four-image system, there are two minimum images and two at saddle points. All of these images are formed at distances from the lens centre of the order of the Einstein radius, which is typically about one arcsecond or about 5--10~kpc at typical lens redshifts of 0.3-1.0.

However, in all of these cases a central image is also formed, corresponding to the maximum in the Fermat surface and generally orders of magnitude closer to the centre of the lensing galaxy. The brightness of this image is related to the shape of the maximum, with sharper maxima giving fainter images. Since the central potential of the galaxy is indeed generally sharply peaked, this gives sharp peaks in the Shapiro delays, which result from passage of light through gravitational fields, and hence relatively faint images. Such central images are therefore very faint, and the detection problem is made worse by the fact that the lensing galaxy is itself both luminous, and liable to cause extinction of the background object.

Although it is difficult to detect, the central image is a potentially valuable probe of matter distributions in the centres of lensing galaxies (Wallington \& Narayan 1993; Rusin \& Ma 2001; Keeton 2003). At these radii -- typically a few tens of parsecs from the centre -- the dark matter contribution is likely to be negligible, and instead the projected mass profile of a galaxy is dominated by the central stellar cusp and a supermassive black hole (SMBH). In nearby galaxies, where the light profile can be observed directly with the Hubble Space Telescope (HST; e.g. Faber et al. 1997), it can be modelled as a broken power law with a break radius that varies widely from galaxy to galaxy; this profile can then be taken as a template for the mass profile. In addition, the mass of the SMBH can be inferred from a strong correlation between $M_{\rm BH}$ and stellar velocity dispersion $\sigma$ (e.g. Ferrarese \& Merritt 2000) which is well established from the cases where $M_{\rm BH}$ can be measured directly from studies of gas disks orbiting the SMBH. The most useful low-redshift comparison galaxies are massive ellipticals of several times $L_*$, since these dominate the strong-lensing cross-section.

\begin{figure*}
\begin{tabular}{cc}
\includegraphics[width=9cm]{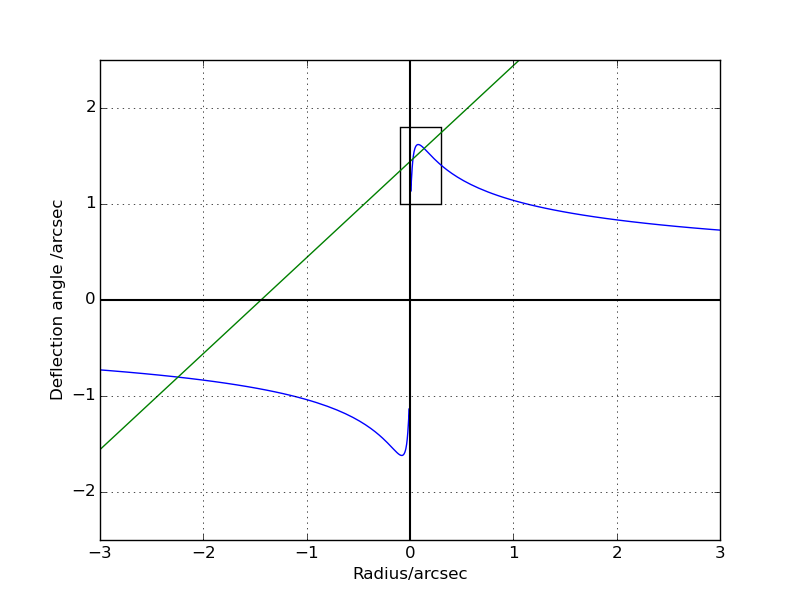}&\includegraphics[width=9cm]{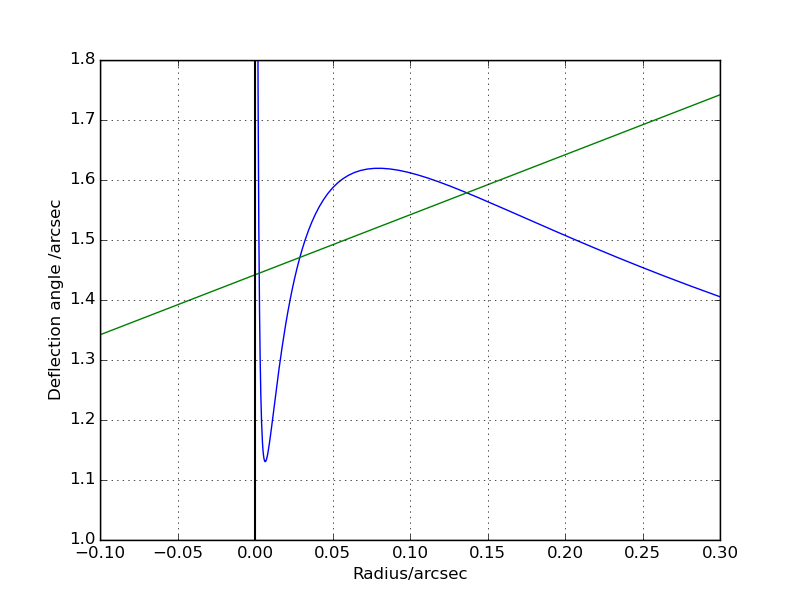}\\

\end{tabular}
\caption{Illustration of the formation of central images; the right-hand plot is an expansion of the part of the left-hand plot shown by the box. A galaxy model and source position have been chosen to roughly reproduce the Einstein radius of the 1030+074 lens galaxy and the ratio of the brightnesses of the images. In both cases, the blue line represents the variation of the gravitationally induced bend angle, $\alpha_{\rm b}$, as a function of angular distance from the centre of the lens galaxy, $\theta$, for a typical galaxy gravitational potential including a black hole. The effect of the black hole is to produce the sharp feature in the middle of the plot, superposed on the S-shaped galaxy feature. The straight green line represents the lens equation $\alpha_{\rm b}-\beta=\theta$, where $\beta$ is the source position with respect to the centre of the galaxy. Lensed images are formed where the curves cross; the line just cuts the top of the curve in the region $r>0$. The magnification of an image is $(\theta/\beta)(d\theta/d\beta)$, so to first order, steep parts of the bend-angle curve yield faint images. Two relatively widely separated bright images form, at around $r=-2.3$ and +0.14 arcsec. The region around the central image can be seen in more detail in the right-hand plot. This shows that an additional image forms at +0.03 arcsec, which is the third image which we are seeking to detect. Two further images form very close to the centre due to the influence of the black hole, which are to all intents and purposes undetectable due to extreme faintness. With this configuration, if the source were slightly further away from the lens, the green line would rise slightly and only one image (at $r=-2.3$ arcsec) would be visible. If the source were slightly closer, the green line would fall, and in this case the third image would approach the centre of the galaxy and become fainter, eventually disappearing once the deflection angle fell below 1.15 arcsec. Note that the image might, however, be visible if the black hole were smaller or absent in this case. Mao, Witt \& Koopmans (2001) give more detailed treatment of all these possibilities.}
\label{geometry}
\end{figure*}

The effects of different mass distributions on lensing properties were studied extensively by Keeton (2003) using the light distributions of Virgo cluster galaxies assembled by Faber et al. (1997). Keeton (2003) found that, for most lensing galaxies and plausible lensing configurations, the factor by which the central image is demagnified ranges from 10$^{-3}$ to 10$^{-5}$. The exception is the case in which the SMBH exceeds a certain critical mass, which itself varies according to the lensing geometry (Fig. \ref{geometry}). In the case where the central image forms close enough to the centre of the lensing galaxy to be affected predominantly by the SMBH potential, the image is demagnified by much larger factors and beyond all realistic prospect of detection. However, for a reasonably large sample of strong lenses, detections should be possible given sufficient sensitivity, and also given sufficiently high resolution to distinguish the central image from the other kiloparsec-scale images of the background object. Two-image lens systems offer much better prospects for detection, since, owing to their geometries, the central image forms further from the SMBH where the potential is less steep; the central image of four-image lenses are generally more heavily demagnified. A particularly favourable case is that of a two-image lens where the source is close to the Einstein radius, leading to a high flux ratio between the two kiloparsec-scale images, a relatively large separation between the central image and the centre of the lensing galaxy, and a less demagnified central image. With luck, and given very high sensitivity, it may be possible to detect a splitting of the central image into two   components (Mao, Witt \& Koopmans 2001; Rusin, Keeton \& Winn 2005) which allows direct measurements of the SMBH mass.

The natural frequency band for detection of central images is the radio, using lens systems with a radio-loud background object. This has three main advantages: high resolution, using modern radio interferometers; relative insensitivity to emission from the lensing galaxy; and relative insensitivity to extinction by passage of the radiation through the intervening lensing galaxy. The insensitivity to lensing galaxy emission and extinction is only relative, however. It is still possible that radio flux may be produced by the lensing galaxy at a low level, but in this case observations over a wide frequency band should allow us to distinguish lensed background radiation from that of the lensing galaxy (Winn et al. 2003, 2004; McKean et al. 2005, 2007; More et al. 2009). More serious is the possibility of scattering, which is known to exist in some lensed radio sources (e.g. Koopmans et al. 2003, Biggs et al. 2004). We discuss this possibility in detail in Appendix A, and suggest that the central image is unlikely to be obliterated by scattering. More recently, Hezaveh, Marshall \& Blandford (2015) have proposed that lens systems found in the submillimetre are also potential central-image systems; in this case, the extended nature of the source also helps in reducing the overall demagnification. Wong, Suyu \& Matsushita (2015) analysed one such lens, SDP81, finding that larger black hole masses ($\log (M_{\rm BH}/M_{\odot})>8.4$) are preferred if the galaxy mass distribution is assumed to follow a cored isothermal profile (see also Tamura et al. 2015).

We do know, however, of one case of a central image in a strong radio-loud lens system, that of the radio lens PMNJ~1632$-$0033 (Winn et al. 2002). This system contains a central image of flux density 1~mJy, strong enough for investigation by the former generation of radio interferometers. Modelling of the system (Winn et al. 2003, 2004) allowed measurements of the central mass profile of the lensing galaxy, in the form of a combination of $M_{\rm BH}$ 
%
and the properties of the central stellar cusp, expressed as the stellar surface density at the central image. These constraints are $M_{\rm BH}<2\times 10^8M_{\odot}$ and $\rho_*>2\times 10^4M_{\odot}{\rm pc}^{-2}$, respectively.

Other observations have historically failed to locate central images. In the northern sky, the major survey for simple two- and four-image lenses is the Cosmic Lens All-Sky Survey (CLASS; Myers et al. 2003, Browne et al. 2003), which found 22 lenses, of which 11 are two-image systems. None have shown the presence of a third central image down to the typical $\sim$50-100~\mujybeam levels achievable using the Very Large (VLA) or extended Multi-Element Remote-Linked Interferometer (MERLIN), both of which had observing bandwidths of $\leq$100~MHz and thus relatively restricted sensitivity. These limits were used by Rusin \& Ma (2001) to derive limits on the central slope, which is likely to be close to isothermal (formally, within 0.2 of the spectral index of the isothermal $\Sigma\propto r^{-1}$, using six lens systems). If the inner slope is steeper than isothermal, the effect is to demagnify the third image to invisibility; however, such a solution is unlikely both on physical grounds and also by comparison with the HST Virgo cluster sample.

Of all the CLASS lenses, the most suitable system for investigation is CLASS~B1030+074. This lens system was discovered by Xanthopoulos et al. (1998) and consists of a highly asymmetric double system (flux ratio $\sim$12 between the two kiloparcsec-scale images) resulting from the lensing of a background quasar of $z=1.535$ by a lens of $z=0.599$ (Fassnacht \& Cohen 1998). The lens appears on optical/IR Hubble Space Telescope images as a faint and slightly irregular galaxy, with a companion (Jackson, Xanthopoulos \& Browne 2000) which is about half an arcsecond, or $\sim3$~kpc, distant. The primary lens is almost certainly an early-type, and is well fit by a de Vaucouleurs profile which is close to spherical ($1-b/a=0.22\pm 0.04$, Leh\'ar et al. 2000). The nature of the companion is more ambiguous, but can be fit by a spherically symmetric exponential disk profile (Leh\'ar et al. 2000). The radio source is bright, with a lensed flux density of $\sim$250~mJy at centimetre radio wavelengths, and the central image is expected to form about 110~mas from the fainter kiloparsec-scale radio image. The radio source is variable, and is expected to show a time delay of about 100 days between variations of the A and B images (Xanthopoulos et al. 1998) but monitoring campaigns have so far failed to yield a convincing delay (G\"urkan et al. 2014, Rumbaugh et al. 2014).

Rusin \& Ma (2001) carried out modelling of this system, with a simple single power-law model and based on the data available at that time, finding a power-law slope constraint $\beta>0.84$, where $\beta=1$ is an isothermal slope. The system was investigated by Zhang et al. (2007) using the High Sensitivity Array (Very Long Baseline Array + Arecibo + Green Bank Telescope + phased VLA). Despite this combination of telescopes, the detection threshold achieved was 180~\mujybeam for any central image, which was not detected. The resulting brightest-to-faintest image ratio of $>1000$ yielded a constraint that either the central mass slope must be within 0.2 of an isothermal slope, or, less plausibly, that the SMBH mass must be $\geq$10 times that implied by the $M_{\rm BH}-\sigma$ relation.

In recent years, both the VLA and MERLIN have undergone extensive upgrades involving the
use of optical fibres to connect the telescopes that make up each interferometer, resulting
in the new Karl G. Jansky Very Large Array (henceforth VLA) and extended MERLIN (e-MERLIN)
instruments, respectively. The sensitivity achievable with these
interferometers is now a factor of 5--10 better than formerly, with sensitivities of
a few $\mu$Jy available in integrations of order 12 hours. For typical lensed image flux
densities of 50-100~mJy, this allows us in principle to compensate for de-magnification factors 
approaching $10^{-5}$. We can thus re-visit the question of central images, with the 
expectation that consistent failure to find central images will now be a surprising result,
which would suggest that more distant lensing galaxies do not resemble their closer elliptical
cousins. As a first step, we have again observed CLASS~B1030+074, but this time with the
VLA and e-MERLIN. In Section 2, we outline the observations and their processing; in 
Section 3, we present the results, and in Section 4 we discuss the implications of the
new and more stringent non-detections. Where necessary, we use a standard flat cosmology
with $H_0$=68~km$\,$s$^{-1}\,$Mpc$^{-1}$ and $\Omega_m$=0.3.

\section{Observations and data reduction}

\subsection{e-MERLIN}

The e-MERLIN array, excluding the 76-m Lovell antenna, observed the gravitational lens CLASS~B1030+074, and calibration sources during 11--15 November 2012. The six remaining antennas which participated in the observations were: Jodrell Bank Mark II (37$\times$25~m); Knockin (25~m); Defford (25-m); Pickmere (25-m); Darnhall (25-m); and Cambridge (32m). Baselines range from 11--217~km, yielding an overall resolution of about 50~mas.

The central frequency was 5.07~GHz with a bandwidth of 0.512~GHz. The observation was divided into four intermediate frequency channels (IFs) with each IF being subdivided into 512 channels of 0.25~MHz width. The absolute flux calibrator was 3C286, while the bandpass (point-like) calibrator and secondary flux calibrator was OQ208 (1407+284). The phase reference source was 1041+0711 with an angular distance from the target of $\sim$1.5$^{\circ}$. The scan strategy consisted of a cycle with seven minutes on target and three minutes on the phase reference source. The phase reference/target cycle consisted of two 13-hour blocks, separated by 11 hours, each block broken up by observations of the flux and bandpass calibrator. 

Data reduction was performed in {\sc aips} \footnote[1]{{\sc aips}: Astronomical Image Processing System, maintained and distributed by NRAO (National Radio Astronomy Observatory), available at http://www.aips.nrao.edu.}. 
Initial inspection of the data revealed radio-frequency interference (RFI) and a strong phase winding across all baselines which contained the Cambridge antenna. This prevented averaging of the data in frequency at this point. All scans within $\sim$4$^{\circ}$ of the horizon, and some within $\sim$7$^{\circ}$, were flagged as they failed to give coherent phase solutions on the phase calibrator. All scans on the bandpass calibrator (and some on the flux calibrator) taken outside of the two phase reference/target cycle blocks were flagged as they produced unsatisfactory complex gain solutions. This is due to time dependent amplitude and phase variations in the array response. The presence of RFI results in an increase in visibility amplitudes. Obvious RFI, as determined by a visual inspection of the visibility amplitudes, associated with specific frequency channels, was flagged. The first 40~s of each scan were flagged to account for all antennas not being on source.


Unknown and variable instrumental delays were determined for the calibration sources and corrections, typically up to 20~nanoseconds, were applied to all sources. 
The complex gains for the calibration sources (OQ208, 1041+0711) were derived from the known flux density of 3C286 (Perley \& Butler 2013) and the point-like nature of the calibration sources. First, phase only solutions were determined, then phase and amplitude solutions were determined, with the amplitudes being scaled to an arbitrary value. The source 3C286 is significantly resolved on e-MERLIN's longer baselines, hence only the shortest baselines in the array (Jodrell Bank Mk2, Pickmere and Darnhall) were used to determine the amplitude scaling from 3C286 to OQ208. 

A separate solution for the flux density in each IF of OQ208 was obtained by bootstrapping from the short baselines
of 3C286, and a fit was performed to set the overall frequency-dependent flux density of OQ208. This was found to 
vary only slightly, from 2.42~Jy at the bottom of the band to 2.46~Jy at the top. The same procedure was used to
find the frequency-dependent flux density of the phase calibrator 1041+0711. RFI which varied both in time and 
frequency, determined by visual inspection, was flagged.



Two 30-minute scans on the bandpass calibrator, taken within each thirteen hour block of observations, were used to determine the frequency dependence of the amplitude and phase response, of each antenna, on a channel by channel basis. The fitted frequency dependent complex gains were applied to the data . The calibration steps to determine the delay solutions and the amplitude scale were repeated on the flagged and bandpass corrected data. 


A self-calibration procedure was undertaken on the phase reference source. Complex gain solutions were determined (phase only) and applied to both the phase reference and target, for successively shorter time intervals (10--0.5 mins). A final phase reference self-calibration cycle produced amplitude and phase solutions over a time interval of 3 minutes. A file was produced, containing only observations of the target, with the derived complex gain solutions applied.

\begin{figure*}
\begin{tabular}{cc}
\includegraphics[width=8.8cm]{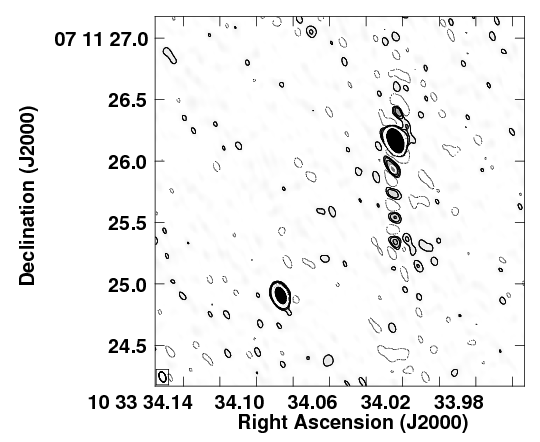}&\includegraphics[width=8.8cm]{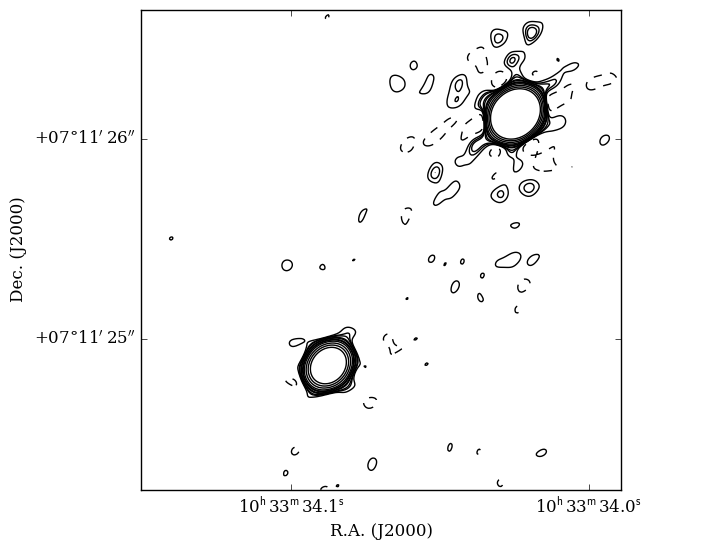}\\
\includegraphics[width=8.8cm]{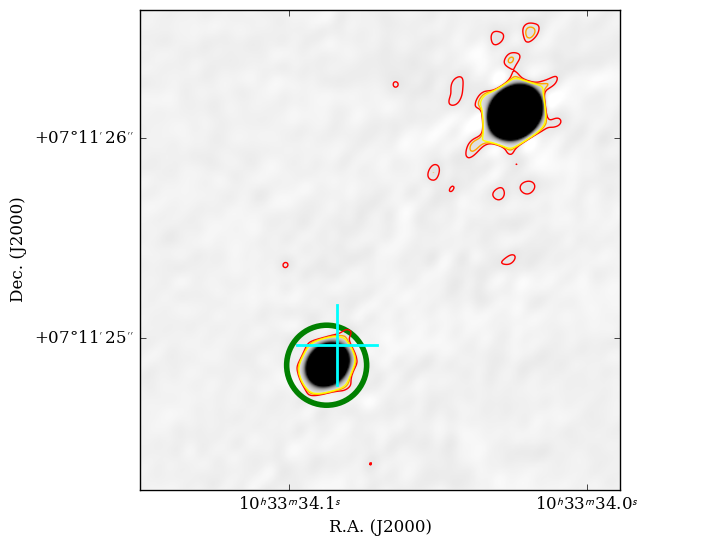}&\includegraphics[width=8.8cm]{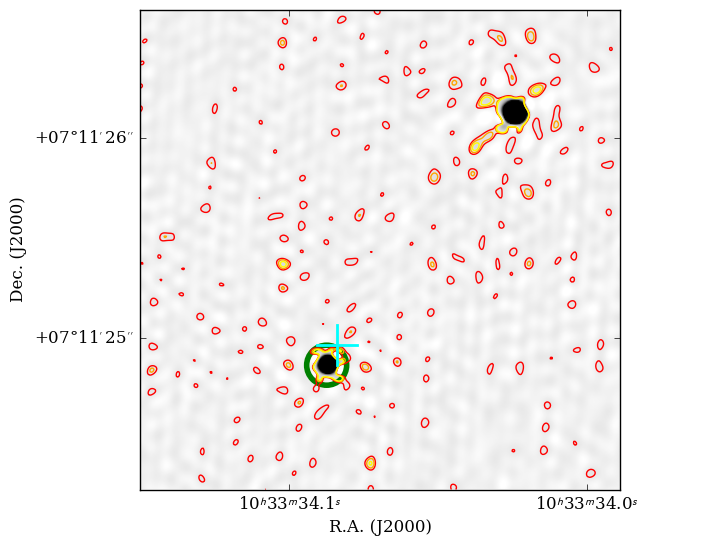}\\
\end{tabular}
\caption{Top: Images of CLASS~B1030+074 with e-MERLIN at 5GHz (left) and with the VLA at 22GHz (right). The e-MERLIN
map is contoured at 50$\mu$Jy/beam $\times$ ($-$1,1,2,4.....32) with a beam of 84$\times$51~mas in PA 21$^{\circ}$, 
and the VLA map at 15$\mu$Jy/beam $\times$(-1,1,2,4...) with optimal (ROBUST=0) weighting and a 100$\times$84-mas beam 
in PA=$-44^{\circ}$. Below: VLA images with different data weightings. In each case the red, orange and yellow 
contours are at 20, 30, and 40~\mujybeam respectively, and the greyscale runs from $-$30$\mu$Jy to 500$\mu$Jy. Each image is labelled with a cyan cross, which represents the position of the lensing galaxy. The central image is expected to form close to this point. The left-hand image has the same beam as the image on the
top right, and the green circle has a radius of 200~mas around the B image. The right-hand image is made with uniform 
weights and a 40-mas beam, slightly smaller than the dirty beam for uniform weighting. Here the green circle has a 
radius of 100~mas.}
\label{images}
\end{figure*}

A map of the target source was produced (Fig. \ref{images}), whose morphology was consistent with previous observations (Zhang
et al. 2007). Areas of the map in which the {\sc clean} algorithm is permitted to assign flux were defined manually, including areas around the bright radio point sources. A cleaned map was then produced, providing a model for self-calibration. Phase only solutions were produced for progressively shorter time intervals (10 to 0.75 mins). After this, amplitude and phase solutions were determined per scan with amplitudes greater than 20~Jy being flagged. The first and last four channels in each IF were flagged, and the remaining channels were averaged to four channels per IF with a bandwidth of 30MHz each. The data were then exported from {\sc aips} into {\sc difmap}\footnote[2]{{\sc difmap}: A program to produce images from visibility data, available at ftp://ftp.astro.caltech.edu/pub/difmap/difmap.html}. Strong, uncorrected phase errors were flagged from the heavily averaged data. A self-calibration cycle was then performed within {\sc difmap}. Firstly, phase only solutions were sought, then amplitude and phase (60 to 1 min). The data were imported into {\sc aips} and a final map was produced with a noise level of 20~\mujybeam .


\subsection{VLA}

The Karl G. Jansky Very Large Array (VLA\footnote[3]{The VLA is operated by the National Radio Astronomy Observatory, 
a facility of the National Science Foundation operated under cooperative agreement by Associated Universities, Inc.}) 
observed the gravitational lens CLASS~B1030+074 over five epochs (see Table 1). Each epoch consisted of two scans of 
CLASS~B1030+074. This was followed by a 159-s scan of the source 3C286, which was used for calibration of the flux 
scale (Baars et al. 1977). Observations were conducted in 16 IFs, each consisting of 64 channels of 2 MHz each with 
the central frequency of the first IF being 20.251 GHz. The bandwidth per IF was 128 MHz, with a total bandwidth of 
2.048~GHz. The relatively high frequency was used, despite a consequent loss of some sensitivity, in order to achieve 
the required resolution; at just under 100~mas, this is the minimum resolution which will allow component B to be
separated from any central component. A total of just over 8 hours of on-source observations were obtained.

\begin{table}
\label{symbols}
\begin{tabular}{@{}lcccccc}
\hline
Epoch & Date & Antenna & No. of& Integration \\
      &      & configuration & antennas & time/min \\ \hline
1 & 19-11-2012 & A & 25  & 100 \\
2 & 09-12-2012 & A  & 24 & 100 \\
3 & 16-12-2012 & A  & 22 & 100 \\
4 & 24-12-2012 & A  & 25 & 100 \\
5 & 29-12-2012 & A  & 25 & 100 \\
\hline
\end{tabular}
\caption{Details of the VLA observations of CLASS~B1030+074.}
\end{table}

The {\sc aips} package was used to process the VLA data with each epoch being reduced separately. The initial scan on the target and the flux
calibrator scan both had the initial 21 seconds flagged to account for all antennas not being on source. Instrumental delays were
calculated and complex gains were corrected for 3C286 to account for its flux and structure, using models available in {\sc aips}. The
corrected visibilities for 3C286 were used to produce an amplitude--frequency calibration, which was applied to CLASS~B1030+074 and
3C286.
The complex gain solutions were re-determined for 3C286, with the visibility amplitude corrections being applied to CLASS~B1030+074.
Complex gain solutions were then determined for CLASS~B1030+074, initially assuming a point source (phase only). A cleaned map was
produced permitting the use of further iterations of a self-calibration procedure. Initially phase only (3--1.25~min) then
amplitude and phase (10--1~min) solutions were produced. Phase stability was generally good, with significant fluctuations
typically on timescales of 10 minutes which allowed good phase solutions to be produced. 

As a result of visual inspection of the bandpass, channels 1 through 5 in IFs 1 and 9 were flagged and channels 1--2 and 63--64 were
flagged in all IFs. Further RFI in several epochs was identified by visual inspection and flagged. The
calibration procedure was then repeated on the flagged data, using the improved source model. Self-calibration was performed 
using {\sc difmap} before reading the data back into {\sc aips}, combining into a single file and imaging using a
range of weighting schemes, from close to natural weighting (which optimizes signal-to-noise at the cost of resolution) to
uniform weighting (which gives better resolution at the cost of signal-to-noise). The maximum noise level attainable in the 
resulting maps (Fig. \ref{images}) is 3~\mujybeam away from the bright images, or slightly higher (about 5--6~\mujybeam)
close to image B. Artifacts at 15--20~\mujybeam levels are present around the bright image A, although these are not a significant
problem for this investigation, as we do not expect the third image to form there.

\section{Modelling and analysis}

\subsection{Constraints provided by the data}

Although the noise levels in the radio maps are well-defined, the constraints provided by the data are not straightforward 
because the central (C) image is expected to form close to the 18-mJy B image, with fainter images forming close to the
centre of the lensing galaxy, 115~mas from B (Jackson, Xanthopoulos \& Browne 2000). This distance is only slightly greater than 
the full width at half maximum (FWHM) of the beam of the naturally-weighted VLA data, and just over twice that of the e-MERLIN
data. After some experimentation, we have based the VLA limits on maps produced with compromise weighting (ROBUST 0 in 
{\sc aips}) but with a restoring beam of 40~mas, just under half of the natural beam. This super-resolution procedure can in 
some circumstances produce spurious structure, or fail to detect actual structure.

Simulations have therefore been undertaken in order to assess our ability to rule out central images of different flux density
in the VLA data, at different distances from the B image. This has been done using artificial sources injected in the data. 
We are particularly interested in the region from 90-115~mas distant from B, as we will see that this is the region in which 
we expect faint third images to form. After artificial sources have been injected into the dataset, the data are then subjected
to the same {\sc difmap} self-calibration and {\sc aips} imaging cycles as the actual map; this tests our ability to recover
sources if they are actually present. In the same way as the actual data, sources close to image B are included in the source
model if they fall close to B.

\begin{figure}
\includegraphics[width=11cm]{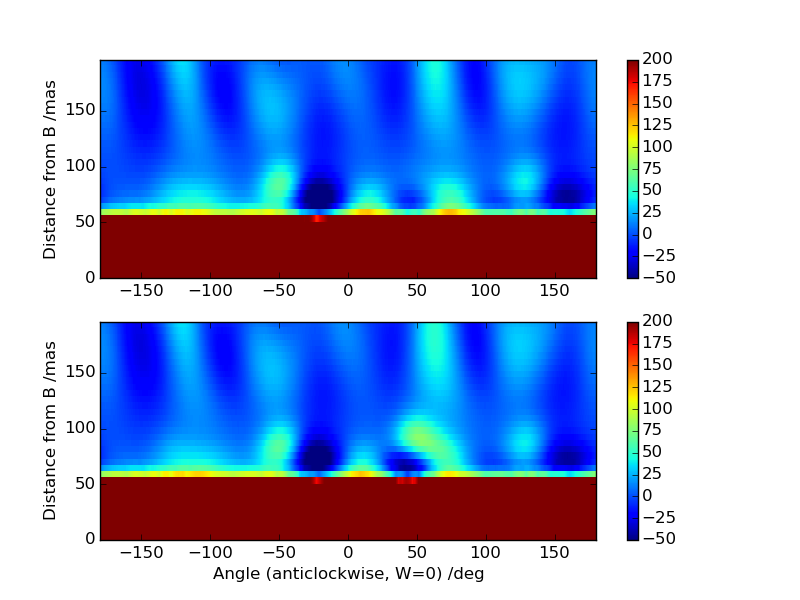}
\caption{Polar plots of the VLA maps close to image B, in the form of flux as a function of angle (anticlockwise from
West) and radius, in milliarcseconds. Image B drowns out any fainter component closer than 50--60~mas, but its influence
decreases rapidly further away than this. In the lower panel, an extra source of 40$\mu$Jy has been inserted into the
data at a distance of 115~mas from B, at a position angle of 53$^{\circ}$. Even after selfcalibration and imaging, it
can clearly be seen, albeit centred at a slightly different point.}
\label{constraints_col}
\end{figure}

Fig.~\ref{constraints_col} shows the region around image B as a plot in polar coordinates, for the data as obtained, and for
the data with a 40-$\mu$Jy point source artificially added at a point 115~mas distant from B, along the line to component A. 
This is approximately the position of the lensing galaxy (Jackson, Xanthopoulos \& Browne 2000) and is the point where a faint
image would be expected to form. The inserted point source is clearly visible, having survived the self-calibration and mapping
process to which the data has been subjected\footnote{The inserted point source will form part of the model, being 
close to the B image, and will therefore have been included in the model used by selfcalibration; however, we stress that the
artificial data has been treated in the same way as the real data for this purpose.}, and is the brightest feature at that 
radius despite appearing slightly closer to B than the point at which it was inserted. More quantitatively, and using the fact 
that we know where we expect any central image to appear, we can express the detection limit at any radius as three times
the rms scatter of the data at that radius from B. This is about 30$\mu$Jy at separations $>$100~mas, but rapidly increases
below this to 70$\mu$Jy at 80~mas and 100$\mu$Jy at 60~mas. The data provide essentially no constraint closer than 50~mas from
the core. However, here the e-MERLIN observations provide information, as the native resolution of the e-MERLIN observations
is twice that of the VLA, and such images would be expected to be much brighter. We have not performed detailed simulations
in the case of the e-MERLIN data, because here the map has not been super-resolved and the dynamic-range requirement is much
less extreme; we have assumed therefore that we would detect any image of flux density $>5\sigma$ which lay more than one beam away.

\subsection{Modelling strategy}

Modelling the mass distributions of gravitational lens systems is a standard problem (e.g. Schneider, Ehlers \&
Falco 1992). A parametrised form is assumed for the mass distribution, and the model parameters are adjusted until
agreement is obtained with the observed positions and flux densities of the lensed images. In the case of the
central image, a non-detection implies a requirement that the model does not predict an image at significantly
greater than the noise level. In general, the positions and flux densities of lensed images give constraints on
the local first and second derivatives of the lens potential, respectively; see Fig. \ref{geometry}, Schneider et al. 
(1992) and Kochanek (1994) for further details.

It is necessary to choose a suitable parametrised distribution for the mass profile. A useful distribution is that
of a broken power law, as this allows for a near-isothermal profile at kiloparsec scales, as inferred by detailed
studies of mass distributions (e.g. Koopmans et al. 2006, Bolton et al. 2008), together with any possible changes
to this profile at low radii. Models similar to this are discussed by Keeton (2003) in his analysis of central images,
and also by e.g. Mu\~noz, Kochanek \& Keeton (2001).

In nearby galaxies, the light distribution can be used as a guide to the mass distribution in the central regions
because the mass profile is expected to be dominated by baryons in this region. Keeton (2003) uses samples of 
nearby early type galaxies (Faber et al 1997, Carollo et al 1997, Carollo \& Stiavelli 1998, Ravindranath 2001),
to determine the inner and outer power law slopes, $\gamma$ and $\beta$, according to a Nuker law (Lauer et al 
1995, Byun et al 1996),

\begin{equation}
I(R) = 2^{\frac{\beta - \gamma}{\alpha}}I_{\rm b} \left(\frac{R}{r_{\rm b}}\right)^{-\gamma}\left[1 + \left(\frac{R}{r_{\rm b}}\right)^{\alpha}\right]^{\frac{\gamma 
- \beta}{\alpha}},    \label{eq:nuker}
\end{equation}

\noindent This broken power law has outer and inner exponents given by $\beta$ and $\gamma$. The brightness scale is set by
$I_{b}$, the surface brightness at the break between the outer and inner regimes. The radius of this break from the
galaxy center is given by $r_{b}$, and its sharpness as $\alpha$. 

Under the assumption that mass follows light, the deflection angle produced by a circular symmetric lens is 
given as (Mu\~noz
et al. 2001; Keeton 2001; Keeton 2003)

\begin{dmath}
\alpha_{\rm gal}(R) = \frac{2^{1 + \frac{(\beta - \gamma)}{\alpha}}}{2 - \gamma}\kappa_{\rm b} r_{\rm b} \left(\frac{R}{r_{\rm b}}\right)^{1 - \gamma} 
\times _2F_1\left[\frac{2 - \gamma}{\alpha},\frac{\beta - \gamma}{\alpha},1 + \frac{2 - \gamma}{\alpha},-\left(\frac{R}{r_{\rm b}}\right)^\alpha\right].    \label{eq:defl}
\end{dmath}

\noindent where $_2F_{1}$ is a hypergeometric function. If we wish to allow for the effect of a black hole, an extra term
$R_E^2/R$ must be added to the deflection, where $R_E$ is the Einstein radius and is proportional to $M_{\rm BH}^{1/2}$ 
(e.g. Schneider, Ehlers \& Falco 1992; Keeton 2003). The mass to light ratio, $\Upsilon$, is contained within the convergence
term, $\kappa_{\rm b}$, such that the surface mass density of the lens at the break radius is $\Upsilon I_b$. Keeton
(2003) gives a table of all parameters for a sample of 73 early-type galaxies, which we use as a basis for the
model constraints presented here. 

We do not consider elliptical models for a number of practical reasons, including
consistency with Keeton (2003). This also avoids further degeneracy in a system with few arcsecond-scale constraints, because the arcsecond-scale images provide too few constraints to marginalise over the extra parameters involved, and gives large increases in computation speed. The main justification for ignoring ellipticity in a single galaxy is that it affects the central image
magnification only very slightly (Keeton 2003). It does, however affect the brightness of the A and B images; a variation
of the shape of the lensing mass distribution from circular to $\epsilon=0.3$ produces a decrease of about 20\% in the 
magnification at the A image position. To first order, this does not make a difference to comparison with the Virgo
galaxies which we undertake in Section 3.4, provided that the ellipticity of the CLASS~B1030+074 lens galaxy is similar to that of the Virgo sample (and
even then, the difference in the effective limit we derive is of order 20\%). In the case of the MCMC analysis of Section 3.5, the assumption
of spherical symmetry has the effect of rendering results slightly conservative. This is because an unmodelled ellipticity 
would cause a sampled lens at the boundary of the ``faint'' (low $\mu_C/\mu_A$) and ``bright'' samples to be erroneously 
included in the ``faint'' sample.

As discussed in Section 1, however, the lensing galaxy has a companion. We can account for the effect that this might have by using the ratio of luminosity of the companion to the primary of Leh\'ar et al. (2000), together with the Faber-Jackson relation, to derive a mass ratio. We then explore the effect that adding this additional mass may have, assuming isothermal models for both. Depending on what constraints are used for images A and B (flux constraints alone, or position and flux constraints), the central image flux density increases by a factor of between 2 and 4, because the influence of the extra mass pushes the central image away from the black hole. This again has the effect of making our inferences in Section 3.4 conservative.

\subsection{Constraints on $r_{\rm b}$ and $\gamma$, assuming an isothermal outer slope}

Neither the VLA 22-GHz map nor the e-MERLIN 5-GHz map shows a third image, to limits that are described in the previous section,
but that reach 30~$\mu$Jy, or about 1/9000 of the flux density of the brightest image, at best. The noise level in Zhang et al.'s (2007) previous VLBI map at 1.7~GHz is 20~\mujybeam,
with constraints at 180$\mu$Jy on the central image. Therefore the current observations represent nearly an order-of-magnitude 
improvement in noise level, assuming a roughly flat radio spectrum, which is consistent both with these observations and
with published photometry. We now explore the implications of this improved result for 
the mass distribution of the lensing galaxy in CLASS~B1030+074.


In addition to the five variables that control the mass distribution of the galaxy ($\alpha$, $\beta$, $\gamma$,
$r_{\rm b}$, $\kappa_{\rm b}$), we also need to consider a sixth variable corresponding to the influence of the central SMBH.
This influence can be parametrised
either in terms of a mass $M_{\rm BH}$ or a velocity dispersion $\sigma$. These may be inferred from each other
by a known empirical relation, in which $M_{\rm BH}/M_{\odot}$ can be fitted approximately by 
$0.282\sigma^{3.75}$ if $\sigma$ is in units of km$\,$s$^{-1}$ (Gebhardt et al. 2000; a relation yielding similar
values for SMBH masses is also given by Ferrarese \& Merritt 2000).  Moreover, the source position $s$ is unknown, but  
can be fitted for any galaxy model, by using the relative arcsecond-scale image positions as constraints.


\begin{figure}
\includegraphics[width=9cm]{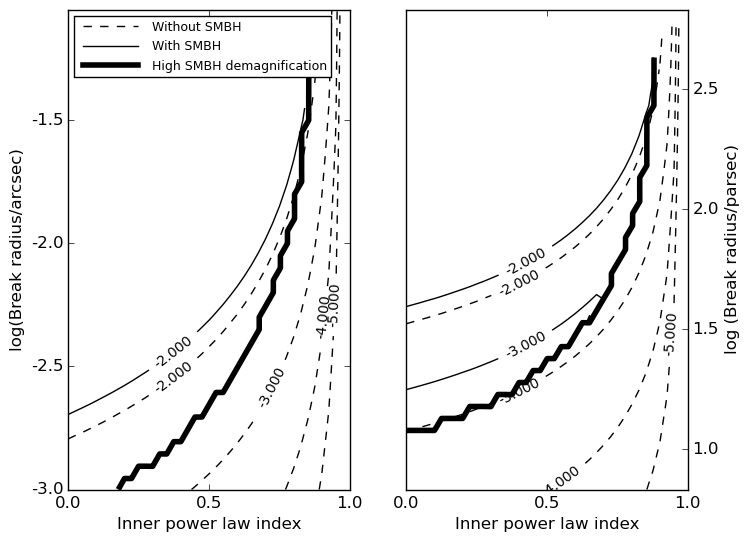}
\caption{Each plot shows the logarithm of the break radius, $r_b$ (in arcseconds on the left and in parsecs on the right, but on the same scale relative to each other) plotted against the inner power law index, $\gamma$, of the Nuker-law which is used to represent the mass profile. The values attached to the contours represent the logarithm of the flux ratio of the central image to the brightest image (i.e. C/A), as a function of these two parameters. The model has in each case been constrained using the observed B/A flux ratio, and the positions of A and B with respect to the galaxy, with the source position and central convergence $\kappa_{\rm b}$ varied for the best fit. Values of $\alpha$ of 1 and 2 have been used for the left- and right-hand plots, respectively; these values of $\alpha$ bracket the large majority of the values in the Keeton (2003) compilation of nearby Virgo galaxies. In each plot we show the case with no black hole (dashed contours) and with a black hole of 2.7$\times 10^{8} M_{\odot}$ (solid contours).  In the latter case, the central image is completely demagnified below and to the right of a locus, represented by the thick line in each plot. In common with other work, an isothermal outer slope ($\beta=1.0$) has been assumed, although this is different from many of the galaxies in the Keeton compilation.}
\label{fig_allowed}
\end{figure}

As a first investigation, and for consistency with previous work (e.g. Rusin \& Ma 2001, Winn et al. 2003, Wong et al. 2015), 
we show the constraints in the plane of break radius vs. inner power law $\gamma$, for the case where the outer power law is 
fixed as isothermal ($\beta=1$; Fig.\ref{fig_allowed}). Most of the galaxies in the Keeton (2003) compilation have outer 
slopes steeper than this, but inner slopes of between 0.0 and 0.7. We perform this with and without a black hole of 
$2.7\times 10^8\,M_{\odot}$ in the model. This SMBH mass has been derived using the observed Einstein radius, and assuming an isothermal profile, to calculate a velocity dispersion $\sigma=233$~km$\,$s$^{-1}$ and hence the black hole mass via the $M-\sigma$ relation. In the simulations, we have used values of $\alpha$=1 and $\alpha$=2 which roughly bracket the values
in the Keeton (2003) compilation. At each grid point we then optimize the value of the convergence, $\kappa_{\rm b}$, and the 
source position, to reproduce the separation and flux ratio of the two bright images, A and B. The main effect of the black 
hole is to demagnify the third image to invisibility for low $r_{\rm b}$, high $\gamma$, or both. The lack of a central image at 
the $C/A \sim 10^{-4}$ level requires either that it is demagnified by a black hole, or that the galaxy parameters $\gamma$ and
$r_{\rm b}$ inhabit a region with close-to-isothermal mass slope and small break radius.  However, we wish to relax the assumptions
we have made and to explore
the parameter space more thoroughly, motivated by two considerations. The first is that the canonical isothermal profile
applies to radii on which the bright images form (typically 5-10~kpc) and results from a ``conspiracy'' between the inner
baryonic profile and the outer, softer dark matter profile; we may expect different results if we are closer to the inner
part of the lensing galaxy's mass profile. The second is that observations exist in the literature of baryonic profiles 
in nearby galaxies, with which it is interesting to compare our constraints in this relatively high-redshift lensing galaxy.

\subsection{Exploring the parameter space: use of Virgo galaxies and A/B flux ratio priors}

We wish to estimate the relevant region of the six-parameter space ($r_{\rm b}$, $\kappa_{\rm b}$,
$\alpha$, $\beta$, $\gamma$, $M_{\rm BH}$) that is allowed by observations. This is a complex exercise, which can be
significantly and usefully simplified by concentrating on parts of this
space that we know to be occupied by nearby galaxies. This will then allow us to test the hypothesis that the mass
distributions of galaxies at $z\sim 0.5$ are systematically different from those of local galaxies. 
Particular areas of this space may be ruled out
by the following considerations:

\begin{enumerate}
\item the galaxy corresponding to this point in parameter space may not have a critical central surface mass density,
and hence not be capable of forming multiple images;
\item the corresponding galaxy may not have the correct Einstein radius to produce the angular separation that we see;
\item the corresponding galaxy may have the correct Einstein radius, but not be able to predict the flux density ratios
of the arcsecond-scale images;
\item the corresponding galaxy may be able to predict the arcsecond-scale positions and flux ratios, but predict a
central image brighter than our limit;
\item the corresponding galaxy may lie in a region of parameter space that is not occupied, or sparsely occupied, by
the observed sample of nearby galaxies tabulated by Keeton (2003).
\end{enumerate}

For our purposes, the first three possibilities are relatively uninteresting, as they can be easily inferred without 
further observations from the fact that we are dealing with a lens system whose separation we know. They do, however, 
apply to a number of members of the nearby galaxy sample,
which are not centrally concentrated enough to form a strong lens system under any circumstances. 

Our approach is therefore to examine galaxies from the Keeton (2003) sample that have approximately the same Einstein radius as CLASS~B1030+074, after correcting for our lens and source redshifts $(z_{\rm l},z_{\rm s})=(0.6,1.0)$ compared to the (0.5,2.0) assumed in the Keeton (2003) models. We then generate simulated galaxies which have the same distribution in the Nuker law parameters ($r_{\rm b}$, $\alpha$, $\beta$, $\gamma$, $k_b$, $M_{\rm BH}$). To do this while taking into account the covariances bewteen various lens model parameters, we use a set of 27 representative Virgo galaxies with a similar Einstein radius (within $\pm$0.2 dex) to infer the underlying probability density function (PDF) from which the galaxies are sampled. Specifically, we use a non-parametric kernel density estimator to build the PDF. For each Virgo galaxy, a six-dimensional kernel, such as a Gaussian, is constructed so that its mean coincides with the observed values. The sum of all the kernels, after being normalised, then estimates the PDF. The width of the kernels is chosen so that the integral of the $\chi^2$ difference between the model and the Virgo samples is minimised. From these we have rejected about 3\% of samples as unphysical, on the grounds that they have $\gamma>1$. For each galaxy, we take the central black hole mass from the $M_{\rm BH}-\sigma$ relation together with the value of $\sigma$ drawn from Table 1 of Keeton (2003). We find the source position by matching the A/B flux ratio to the ratio of 11.7$\pm$1.2 seen in 1030+074. This is important, because the likelihood of seeing a central image is a strong function of flux ratio. We also fit the observed ratio of lengths from the lensing galaxy to the A and B images. The combined procedure usually leads to a separation of the B image from the galaxy that is inconsistent with the observed value (Jackson, Xanthopoulos \& Browne 2000), so we scale each simulated lens so that the predicted separation of the galaxy and B agrees with the observed value, in the process scaling the offset of the central image from the lens galaxy. We then calculate the central-image positions and fluxes associated with this artificial sample.

\begin{figure*}
\begin{tabular}{cc}
\includegraphics[width=8cm]{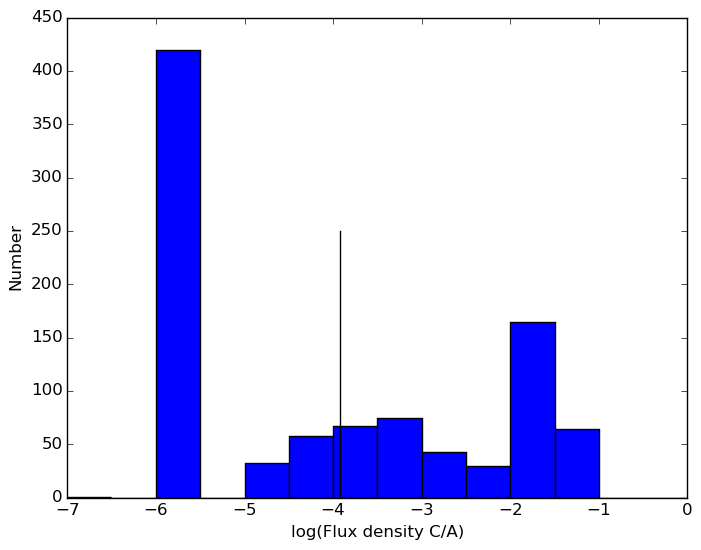}&\includegraphics[width=8cm]{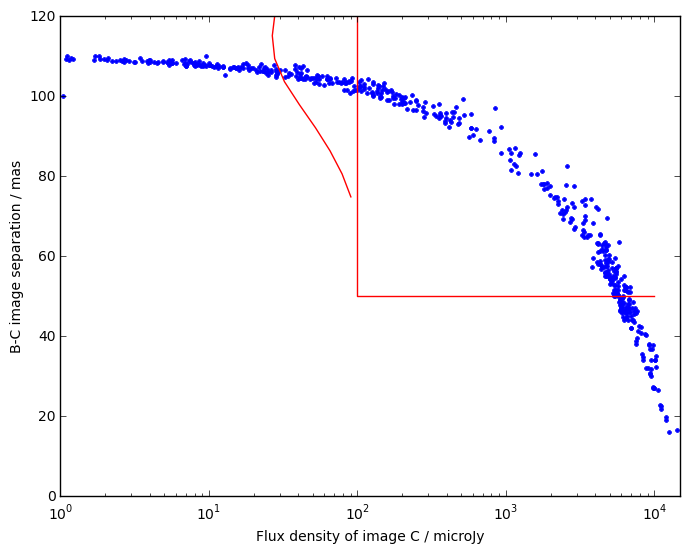}\\
\end{tabular}
\caption{Left: Histogram of the logarithm of the ratio of the flux of the central image to that of the brightest (A) lensed image in the simulated sample. Simulated objects in which the black hole renders the central image close to infinitely demagnified are reported as -6. The limit we have derived from the radio observations ($-$3.9, shown as a vertical black line) is nearly an order of magnitude more sensitive than the previous best limit, allowing us to rule out all but 10\% of cases in which the central image is just below our limit, together with about 45\% of cases in which the central image is essentially erased by the gravitational action of a central SMBH. Right: for 1000 objects of the simulated sample, the expected flux density of the central image against its distance from the B image. In general, brighter central images form closer to the B image. The two lines represent the limit from the VLA and e-MERLIN observations, respectively, with the e-MERLIN data contributing to rule out closer images despite its lower sensitivity. The lines have been determined as described in the text; in the case of the VLA, from simulations, and for e-MERLIN on the assumption that sources of 5 times the r.m.s. would be detected more than 1 beam away. The very few bright images outside either constraint are inconsistent with VLBI data (Zhang et al. 2007). Only 55\% of the 1000 simulated sources appear on this diagram; the others have central images which are extremely demagnified.}
\label{magsims}
\end{figure*}

The results are shown in Fig.~\ref{magsims}. We find that in about 45\% of cases, the SMBH results in the central image of the simulated objects being unobservable (with demagnifications of typically 10-20 orders of magnitude). In the majority of cases, however, the third image is expected to contain somewhere between $10^{-4}$ and $10^{-1}$ of the flux of the brightest image. In a few cases, less than 10\% of the total simulations, the third image is anomalously bright; such cases correspond to situations similar to that illustrated in Fig.~\ref{geometry}. In previous work, we were not able to exclude much of the main part of this histogram, because the central image limit was not sensitive enough. Our current limit, approximately $10^{-3.9}$, excludes all but 15--20\% of models in which the central SMBH does not heavily demagnify the central image, the remaining such models being still consistent with the data. Two conclusions follow from this observation, if we assume that galaxies at typical lens redshifts are similar to local ellipticals. Firstly, these observations have almost reached the useful limit of depth in this particular object, and we would predict that it is now more likely than not that no further level of observational effort on this object will now yield a central image detection. Secondly, on the basis of these models we would expect to see a non-detection at this level about half of the time. Observations of 5--10 more objects at this extreme dynamic range should therefore provide a definitive test of the hypothesis that distant elliptical galaxies have similar mass distributions to local ones.

We have performed this analysis by matching, for each simulated galaxy, the flux and distance ratios of the A and B 
images in order to find the source position. We could, instead, have matched the positions of the A and B images with 
respect to the lensing galaxy, ignoring the fluxes; if the galaxy is isothermal and singular this gives the same result. 
For the general shape of the bend angle diagrams (e.g. Fig. 1) in the galaxies we are considering, however, the central 
images are  typically brighter than those derived from fitting the ratios. The results we obtain are in fact 
slightly conservative, for this reason and also because, as previously pointed out, we have ignored the companion in the lens model.

\begin{figure*}
\includegraphics[width=16cm]{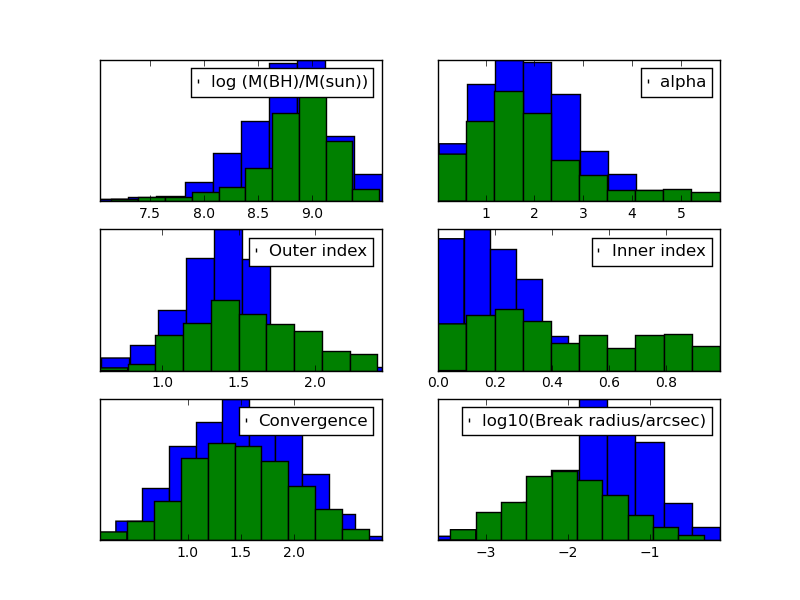}
\caption{Histogram of properties of the sample of galaxies which give observable central images (blue) and highly
demagnified central images (green). In general, the latter sample generally, though not universally, consist of
galaxies which contain higher-mass black holes. However, the main difference is that galaxies in the latter sample
have lower central convergence and break radii of the lens galaxy potential, excluding the black hole.}
\label{maghist}
\end{figure*}

In Fig. \ref{maghist} we compare the properties of the galaxies that produce observable and unobservable central images.
Although the galaxies that produce unobservable central images have a slight tendency to contain more massive black holes,
the dominant effect is the tendency of such galaxies to have lower central convergence, higher inner power law slope or both.
For a given black hole mass and A/B image flux ratio, the observability of a central image is thus mainly controlled
by the properties of the centre of the lens galaxy, at least for the range of black hole masses considered here. It is
thus dangerous to interpret a non-detection of a central image simply in terms of a black hole mass limit, although 
given any particular galaxy model, a
large enough black hole mass will eventually remove the central image produced by any lens galaxy.


\subsection{Exploring the parameter space: using only measured parameters as priors}

\begin{figure*}
\begin{tabular}{c}
\includegraphics[width=11cm]{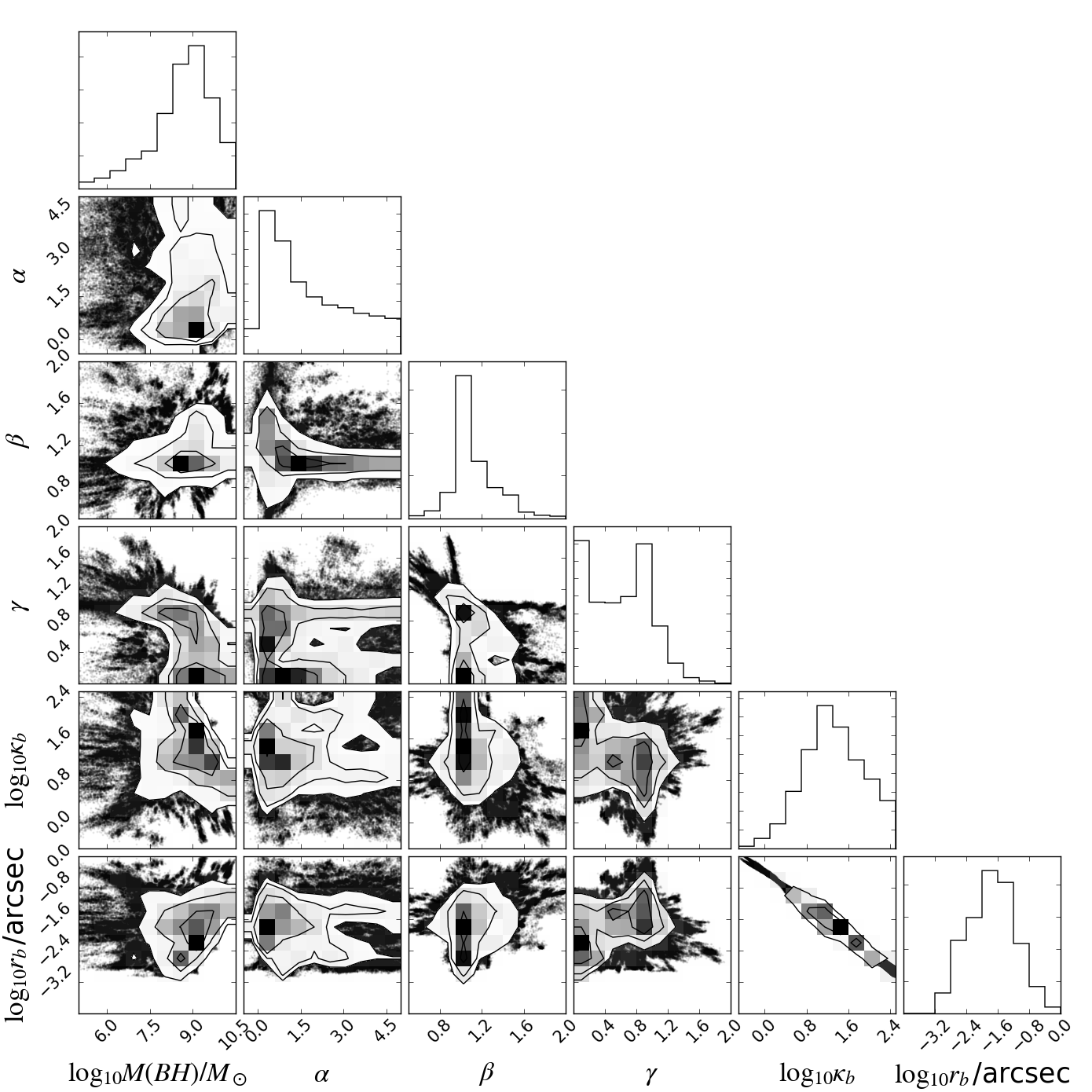}\\
\includegraphics[width=11cm]{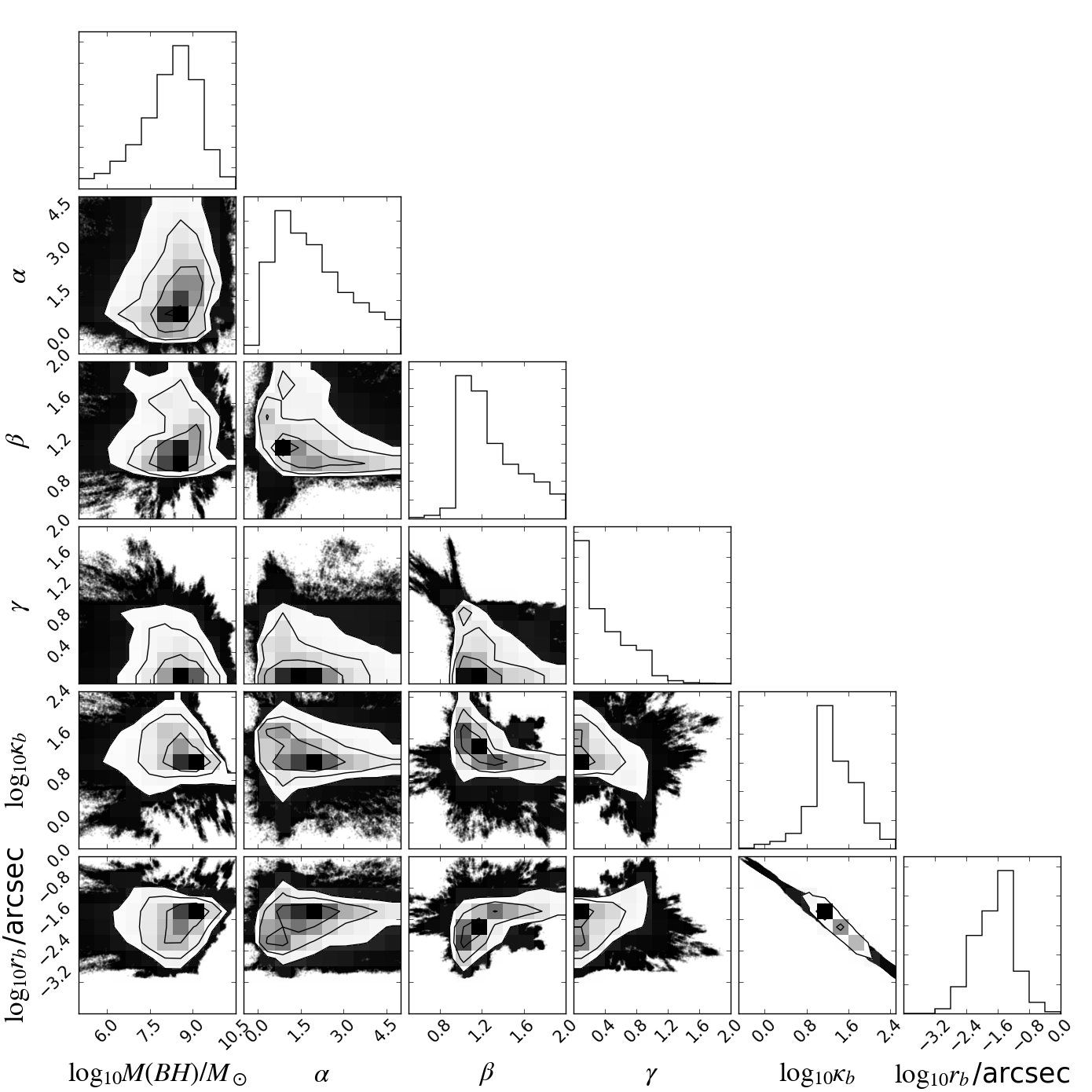}\\
\end{tabular}
\caption{Results of modelling with galaxy parameters unconstrained by comparison with Virgo galaxies, but requiring
only that the Einstein radius, A/B flux ratio and A/B separation ratio be reproduced, and with a prior of monotonic
decrease of density with radius ($\gamma\geq 0$). Top: plots of the MCMC run, plotting
only those samples which are allowed by our new constraint on the flux density of the central image. Bottom: all samples.}
\label{mcmc_magsims}
\end{figure*}

In the previous subsection, we required only that the observations should reproduce the observed A/B flux ratio, but did
not also demand compliance with the measured Einstein radius and image separation ratio (i.e. the ratio of the separations
of A and B from the galaxy. We now therefore attempt to reproduce the Einstein radius (1567$\pm$5~mas), and the flux and
separation ratios (11.7$\pm$1.2 and 12.5$\pm$1.1, e.g. Jackson, Xanthopoulos \& Browne 2000). We perform a Markov Chain
Monte-Carlo procedure using the publicly available {\sc emcee} code (Foreman-Mackey et al. 2013). The initial positions
of the MCMC walkers are determined by starting from the galaxy parameters of each Virgo galaxy, and optimising to the
point at which $\chi^2=0$ (since the fit is underdetermined, having seven free parameters in the form of six specifying 
the galaxy profile and an unknown source position). This MCMC procedure allows very much more freedom, and includes a
number of galaxy profiles that are probably unphysical (Fig.~\ref{mcmc_magsims}). We do, however, apply a prior on the
inner power law slope of $\gamma\geq 0$, which ensures that the density decreases monotonically from the centre of the
lensing galaxy. For each point in the MCMC, we optimise
the source position for the best fit to the observational constraints provided by the A and B images, but do not attempt
to use the third image constraint. We then separate each MCMC accumulated sample into a ``faint'' sample -- whose third
images are consistent with our observational constraint -- and a ``bright'' sample, whose third images are not consistent.

\begin{figure*}
\begin{tabular}{c}
\includegraphics[width=11cm]{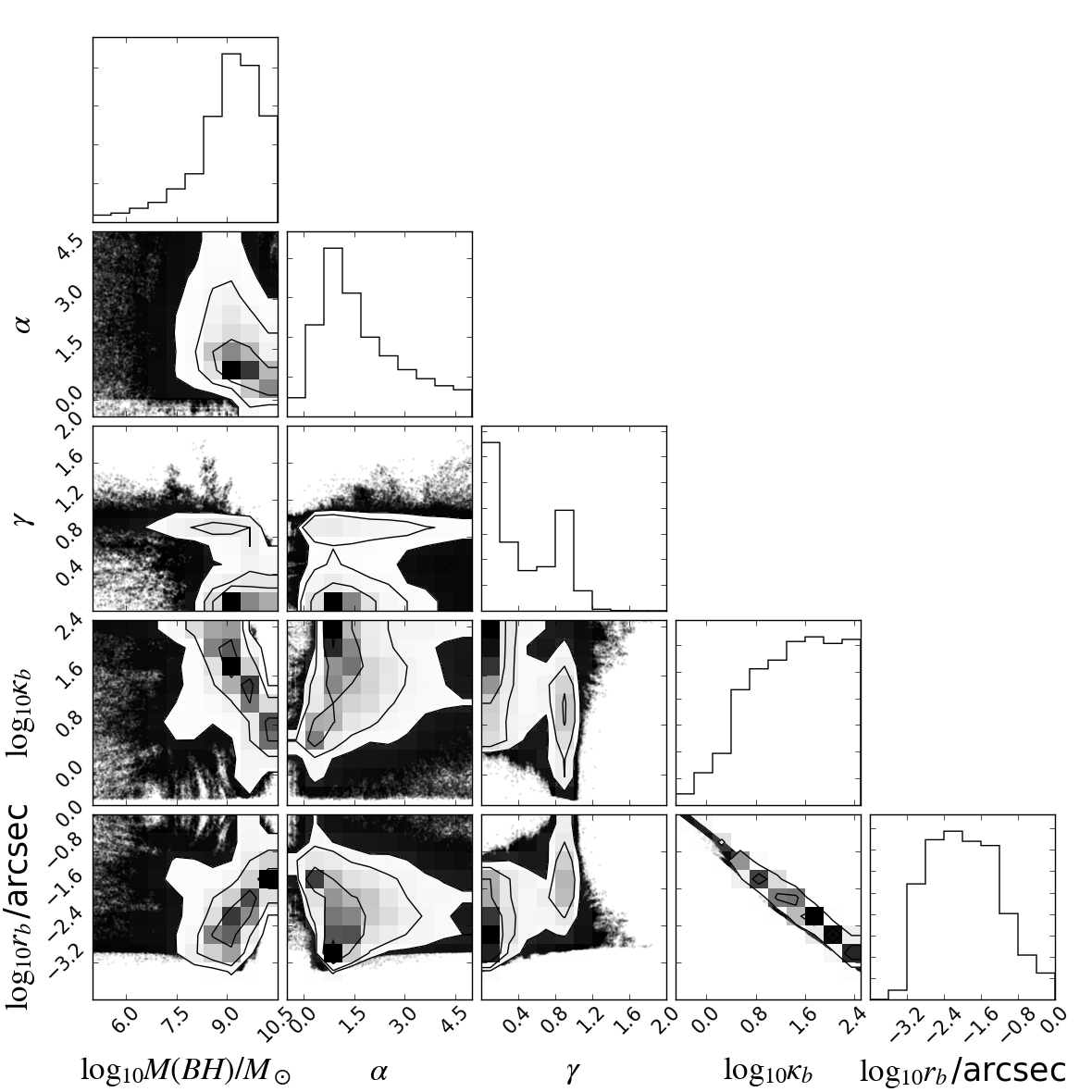}\\
\includegraphics[width=11cm]{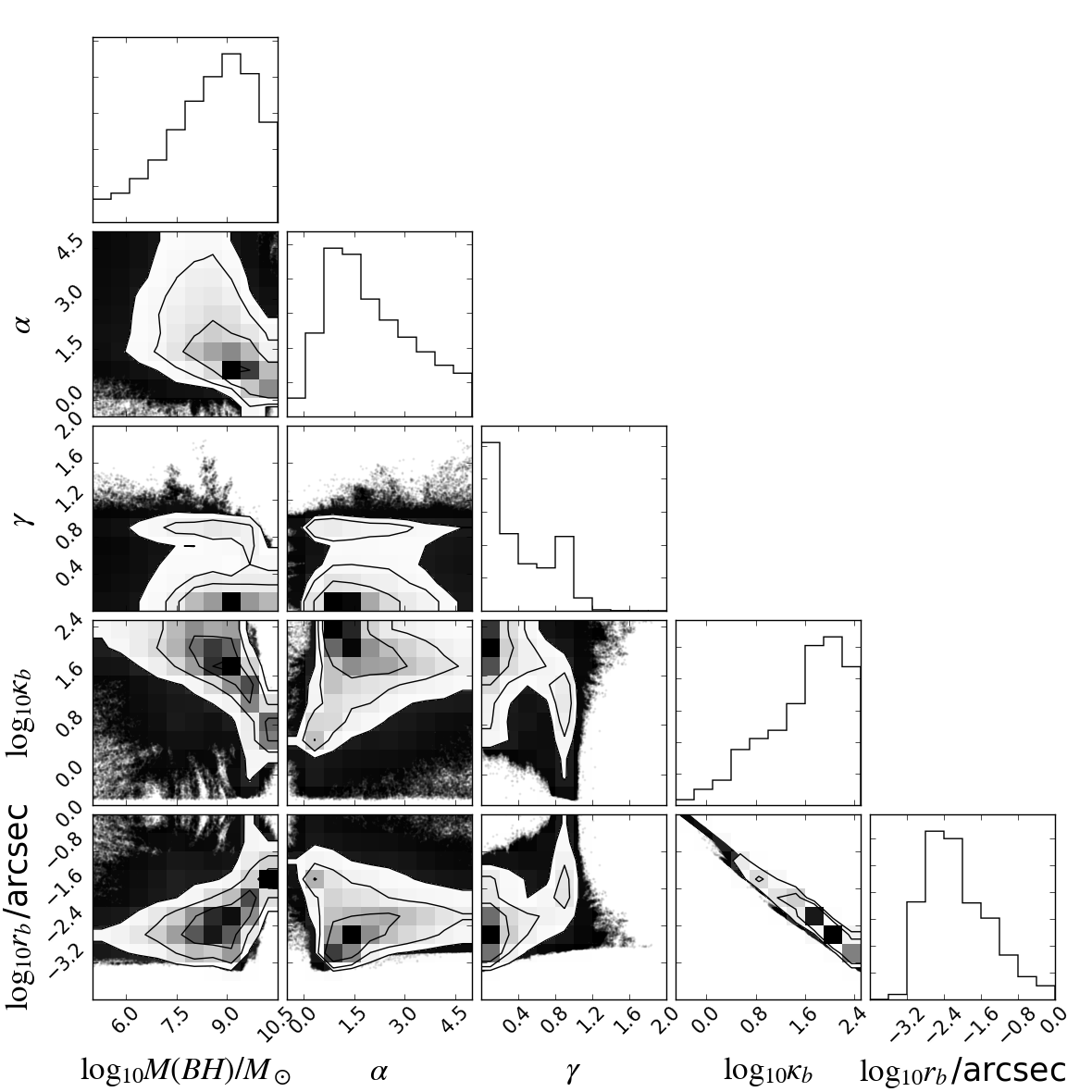}\\
\end{tabular}
\caption{Results of modelling with galaxy parameters, including an outer isothermal slope ($\beta$=1) and a prior of
monotonic decrease in density with radius ($\gamma>0$). The modelling
is unconstrained by comparison with Virgo galaxies, but requiring
only that the Einstein radius, A/B flux ratio and A/B separation ratio be reproduced. Top: plots of the MCMC run, plotting
only those samples which are allowed by our new constraint on the flux density of the central image. Bottom: all samples.}
\label{mcmc_magsims_iso}
\end{figure*}

\begin{table*}
\begin{tabular}{lcccccc}
& \multicolumn{3}{c}{$\beta$ free}&\multicolumn{3}{c}{Isothermal}\\
& All & Bright & Faint & All & Bright & Faint \\
$\log_{10}M_{\rm BH}$ & $8.08^{+0.88}_{-1.95}$ & $7.98^{+0.81}_{-1.92}$ & $8.35^{+1.01}_{-2.08}$ & $7.64^{+1.56}_{-2.57}$ & $7.07^{+1.13}_{-2.24}$ & $8.02^{+1.44}_{-2.83}$ \\
$\alpha$ & $2.19^{+4.16}_{-1.39}$ & $2.06^{+3.48}_{-1.16}$ & $2.75^{+4.94}_{-2.23}$ & $3.35^{+7.97}_{-2.37}$ & $3.32^{+6.38}_{-2.11}$ & $3.37^{+8.69}_{-2.47}$ \\
$\beta$ & $1.19^{+0.52}_{-0.16}$ & $1.28^{+0.51}_{-0.21}$ & $1.04^{+0.25}_{-0.11}$ & $1.00^{+0.00}_{-0.00}$ & $1.00^{+0.00}_{-0.00}$ & $1.00^{+0.00}_{-0.00}$ \\
$\gamma$ & $0.36^{+0.51}_{-0.27}$ & $0.27^{+0.40}_{-0.20}$ & $0.76^{+0.26}_{-0.60}$ & $0.42^{+0.53}_{-0.33}$ & $0.30^{+0.43}_{-0.23}$ & $0.50^{+0.48}_{-0.41}$ \\
$\log_{10}\kappa$ & $1.22^{+0.57}_{-0.38}$ & $1.21^{+0.44}_{-0.28}$ & $1.27^{+1.20}_{-0.72}$ & $2.01^{+1.08}_{-1.02}$ & $1.94^{+0.40}_{-0.38}$ & $2.13^{+1.31}_{-1.37}$ \\ 
$\log_{10}r_b$/arcsec & $-1.55^{+0.43}_{-0.68}$ & $-1.51^{+0.36}_{-0.57}$ & $-1.70^{+0.77}_{-1.20}$ & $-2.51^{+1.04}_{-1.12}$ & $-2.41^{+0.42}_{-0.48}$ & $-2.65^{+1.39}_{-1.28}$ \\
\end{tabular}
\caption{Median values of the fitted galaxy parameters in the MCMC simulations (see Figs.~\ref{mcmc_magsims} and \ref{mcmc_magsims_iso}, with scatters formed by the 16th and 84th percentiles of the distributions. Statistics are given for the free-index simulation (Fig.~\ref{mcmc_magsims}) and for the isothermal simulation (Fig.~\ref{mcmc_magsims_iso}). For each simulation, we give the statistic for the whole sample, the portion of the sample brighter than our detection limit, and the portion fainter than our detection limit.}
\label{mcmctable}
\end{table*}

The results of the MCMC runs (Figs.~\ref{mcmc_magsims} and \ref{mcmc_magsims_iso} and Table~\ref{mcmctable}) are interesting, and illustrate the complicated nature of the parameter space. The one
very clear relationship, persisting through all investigations, is an anticorrelation between the overall convergence, 
$\kappa_{\rm b}$, and the break radius $r_{\rm b}$. In the first MCMC run we allow the outer slope $\beta$ to vary (Fig.~\ref{mcmc_magsims}) 
and investigate the differences between the corresponding ``faint'' and ``bright'' samples. As expected, we find that 
systems with faint third images have larger SMBH masses and lower break
radii, although the separation is not enormous; for example, the difference in SMBH masses is about 0.5~dex and there is
a large degree of overlap in the two distributions. The inner power law slope, $\gamma$, is generally shallower in systems with 
bright third images, although again there is considerable overlap in the distribution of this parameter between systems
which produce bright and faint third images. However, no system with significant third-image flux may have an inner
power law slope which is steeper than isothermal. In general, the interplay of the SMBH with the parameters controlling the
shape of the galaxy mass distribution conspire to allow a substantial amount of degeneracy in the manner in which any
particular third-image flux can be produced.

We next fix the outer slope to isothermal ($\beta=1$), again with the caveat that the radius at which this fixes the
slope is in general considerably less than the radius at which the slope has been reliably measured to be isothermal in other lens systems.
Here, we do obtain a clear separation in black hole masses using our third-image constraint (Fig.~\ref{mcmc_magsims_iso}). The corresponding ``faint''
and ``bright'' samples are now separated, with the systems producing faint third images clustering between 
$10^{8.5}$ and $10^9~M_{\odot}$. The systems producing bright third images, by contrast, cut off at this point. For 
this value of $\beta$ we do not observe any separation in break radius between ``faint'' and ``bright'' samples, the main
difference between the distributions being that the systems producing faint third images have a much wider range of
likely $r_{\rm b}$ values. We also observe a bimodal distribution of the inner power law slope $\gamma$ within the samples, with
many samples clustering close to the $\gamma=0$ cutoff and the remainder close to isothermal. This effect is a result of
the requirement that we fit the A and B image flux ratios and positions, and would be different for any other lens system.

\section{Conclusions}

We have reported the initial results of a programme aimed at an eventual systematic VLA/e-MERLIN survey for central images in
radio-loud gravitational lens systems, using the new and greatly increased sensitivities of modern radio 
interferometers. We have achieved a high dynamic range in the images corresponding to a detection limit of nearly 10000:1 in 
the ratio of the central image to the brightest arcsecond-scale image, a factor of about 10 better than the previous limit. 
The unprecedentedly tight constraint on the central image has been used in a number of different analyses, in conjunction
with a double power-law model motivated by studies of the light distribution in nearby elliptical galaxies. Based on this
modelling, we find the following results:

\begin{itemize}
\item For simple models with a fixed isothermal outer power law slope, a SMBH is required to explain our non-detection
of the central image, unless the break radius between the outer and inner power laws is extremely small (a few parsecs)
or the inner power law slope is within 0.1 of isothermal. More realistically, if we assume a SMBH of a mass only slightly
greater than that implied by the $M_{\rm BH}:\sigma$ relation, plausible combinations of the break radius, inner power 
law slope and the $\alpha$ parameter, which
controls the smoothness of the break, can be found. The assumption of a fixed isothermal
outer power law slope is frequently made in the literature, although it does originate in measurements made on scales
rather greater than those being considered in studies of the central images.

\item We have also confronted the 
observations with models of elliptical galaxy mass distributions, assuming that the central regions of these can be fairly 
represented by light distributions from HST observations of nearby elliptical galaxies. Under this assumption, we find that 
the chance of suppressing the third image without the action of an SMBH is now relatively small. About 10\% of simulations, using
parameters similar to those which describe nearby galaxies, give images fainter than our detection limit, with the majority
of these being only slightly fainter. A further 45\% are very much more demagnified by the action of a SMBH. In order for 
the SMBH to effect this demagnification, we find that lens galaxy mass distributions containing small break radii (less than 
about 15~mas, corresponding to about 70~pc) are preferred.

\item Finally, we have for the first time in a single lens system investigated the more complicated problem of the interplay
of all of the lensing-galaxy mass parameters using an MCMC process. Full investigation of the parameter space reveals a complex
dependence of the observed quantities on a combination between the SMBH mass and the parameters describing the lensing galaxy 
mass. If simplifying assumptions are made -- essentially if the outer power law slope is required to be isothermal -- then a 
separation in SMBH mass appears between systems which produce central images consistent with our upper limit, and those which 
do not. The split occurs at about 3$\times$10$^8$~$M_{\odot}$, close to the SMBH mass implied by the $M:\sigma$ relation;
for MCMC samples which are still consistent with our detection limit, the majority of samples have a SMBH considerably more
massive than that implied by the $M:\sigma$ relation. The analysis presented here is specific to the lens system 
CLASS~B1030+074, and less good constraints on lens galaxy parameters may be expected in systems where the flux ratio of the 
two bright images is smaller.

\item A number of more complicated relations between the lens galaxy mass parameters are apparent, most of which are due to
the details of the fits to the positions and flux densities of the A and B images. Further investigations of a number of 
lens systems will be needed in order to integrate over the peculiarities of each individual system and arrive at more
general conclusions.

\item We suggest (Appendix A) that scattering and free-free
absorption are unlikely to be responsible for the non-detection of the central image. 
\end{itemize}

Finally, the observations and analysis 
we present here suggest that observations of a modest further number of targets will either detect central images, or imply
that mass distributions of distant elliptical galaxies which produce gravitational lensing are different from the
mass distributions of elliptical galaxies in the local Universe.

\section*{Acknowledgements}

We thank an anonymous referee for useful comments on this paper. Jonathan Quinn acknowledges receipt of an STFC studentship. e-MERLIN is a National Facility operated by the University of Manchester
at Jodrell Bank Observatory on behalf of the UK Science and Technology Facilities Council. The Karl G. Jansky VLA is operated by
the US National Radio Astronomy Observatory (NRAO). NRAO is a facility of the U.S. National Science Foundation operated under
cooperative agreement by Associated Universities Inc. GDZ acknowledges financial support from ASI/INAF agreement n. 2014-024-R.0

\section*{References}

\noindent Baars J.W.M., Genzel, R., Pauliny-Toth, I.I.K., Witzel, A., 1977, A\&A, 61, 99.

\noindent Biggs, A.D., et al., 2004, MNRAS, 350, 949.

\noindent Bolton, A.S., et al., 2008, ApJ, 682, 964.

\noindent Browne, I.W.A., et al., 2003, MNRAS, 341, 13.

\noindent Byun, Y.-I., et al., 2006, AJ, 111, 1889

\noindent Carollo, C.M., Stiavelli, M., 1998, AJ, 115, 2306.

\noindent Combes, F., Wiklind, T., 1998, A\&A, 334L, 81.

\noindent Cordes, J.M., Lazio, T.J.W., 2003, astro, ph, 1598.

\noindent Faber, S.M., et al., 1997, AJ, 114, 1771.

\noindent Fassnacht, C.D., Cohen, J.G., 1998, AJ, 115, 377.

\noindent Ferrarese, L., Merritt, D., 2000, ApJ, 539L, 9.

\noindent Gebhardt, K., et al., 2000, ApJ, 539L, 13.

\noindent G\"urkan G., Jackson N, Koopmans L.V.E., Fassnacht C.D., Berciano Alba A., 2014, MNRAS, 441, 127

\noindent Hezaveh Y., Marshall P.J., Blandford R.D., 2015, ApJ, 799, L22.

\noindent Jackson, N., Xanthopoulos, E., Browne, I.W.A., 2000, MNRAS, 311, 389.

\noindent Keeton, C.R., 2001, astro, ph, 2341.

\noindent Keeton, C.R., 2003, ApJ, 582, 17.

\noindent Kochanek C.S., 2004, in Schneider P., Kochanek C.S., Wambsganss J.,
Proc. Saas-Fee School ``Gravitational Lensing: Strong, Weak and Micro'', eds. Meylan G. et al., Springer, Berlin.

\noindent Koopmans, L.V.E., Treu, T., Bolton, A.S., Burles, S., Moustakas, L.A., 2006, ApJ, 649, 599.

\noindent Koopmans, L.V.E., et al., 2003, ApJ, 595, 712.

\noindent Lauer, T.R., et al., 1995, AJ, 110, 2622.

\noindent Leh\'ar, J., et al., 2000, ApJ, 536, 584.

\noindent Mao, S., Witt, H.J., Koopmans, L.V.E., 2001, MNRAS, 323, 301.

\noindent Marlow, D.R., et al., 1999, AJ, 118, 654.

\noindent McKean, J., et al., 2005, MNRAS, 356, 1009.

\noindent McKean, J., et al., 2007, MNRAS, 378, 109.

\noindent Mittal, R., Porcas, R., Wucknitz, O., 2007, A\&A, 465, 405.

\noindent Mittal, R., Porcas, R., Wucknitz, O., Biggs, A., Browne, I., 2006, A\&A, 447, 515.

\noindent More, A., McKean, J., More, S., Porcas, R.W., Koopmans, L.V.E., Garrett, M.A., 2009, MNRAS, 394, 174.

\noindent Mu\~noz J., Kochanek C.S., Keeton C., 2001, ApJ, 558, 657

\noindent Myers, S.T., et al., 2003, MNRAS, 341, 1.

\noindent Patnaik, A.R., Browne, I.W.A., Walsh, D., Chaffee, F.H., Foltz, C.B., 1992, MNRAS, 259P, 1.

\noindent Perley, R.A., Butler, B.J., 2013, ApJS, 204, 19.

\noindent Phillips, P.M., et al., 2000, MNRAS, 319L, 7.

\noindent Ravindranath, S., Ho, L.C., Peng, C.Y., Filippenko, A.V., Sargent, W.L.W., 2001, AJ, 122, 653.

\noindent Rickett, B.J., 1977, ARA\&A, 15, 479.

\noindent Rumbaugh, N., Fassnacht, C.D., McKean, J.P., Koopmans, L.V.E., Auger, M.W., Suyu, S.H., 2014. MNRAS, 450, 1042

\noindent Rusin, D., Ma, C., 2001, ApJ, 549L, 33.

\noindent Rusin, D., Keeton, C.R., Winn, J., 2005, ApJ, 627, L93.

\noindent Scalo, J., Elmegreen, B.G., 2004, ARA\&A, 42, 275.

\noindent Schneider, P., Ehlers, J., Falco, E.E., 1992, Gravitational Lenses, Springer-Verlag, Berlin.

\noindent Schneider, P., Weiss, A., 1992, A\&A, 260, 1.

\noindent Tamura, Y., Oguri, M., Iono, D., Hatsukade, B., Matsuda, Y., Hayashi, M., 2015, PASJ, 67, 4, 727.

\noindent Taylor, J.H., Cordes, J.M., 1993, ApJ, 411, 674.

\noindent van Langevelde H., Frail, D.A., Cordes, J.M., Diamond, P.J., 1992, ApJ, 396, 686

\noindent Walker, M.A., 1998, MNRAS, 294, 307.

\noindent Wallington, S., Narayan, R., 1993, ApJ, 403, 517.

\noindent Wiklind, T., Combes, F., 1995, A\&A, 299, 382.

\noindent Winn, J.N., Rusin, D., Kochanek, C.S., 2003, ApJ, 587, 80.

\noindent Winn, J.N., Rusin, D., Kochanek, C.S., 2004, Natur, 427, 613.

\noindent Winn, J.N., et al., 2002, AJ, 123, 10.

\noindent Wong, K.C., Suyu, S.H., Matsushita, S., 2015, astro-ph/1503.05558

\noindent Xanthopoulos, E., et al., 1998, MNRAS, 300, 649.

\noindent Zhang, M., Jackson, N., Porcas, R.W., Browne, I.W.A., 2007, MNRAS, 377, 1623.

\section*{APPENDIX A: Propagation effects}

By definition, the radiation from background sources in gravitational lens systems propagates through the medium of the
lensing galaxy. At radio wavelengths, the main resulting effect is scatter broadening of the image, associated with passage
through inhomogeneous ionized media (see e.g. Rickett 1977; Scalo \& Elmegreen 2004). The most spectacular
example of this is in the lens system CLASS~B0218+357 (Patnaik et al. 1992) in which one of the lensed images is seen
through a giant molecular cloud associated with the spiral lens (Wiklind \& Combes 1995; Combes \& Wiklind 1998) and
appears to be scatter-broadened as a result (Mittal et al. 2006; Mittal, Porcas \& Wucknitz 2007). Evidence for scatter
broadening is also seen in other radio lenses, notably CLASS~B0128+437 (Phillips et al. 2000; Biggs et al. 2004),
CLASS~B1933+503 (Marlow et al. 1999) and PMNJ1838$-$3427 (Winn et al. 2004). In many of these cases the lensing galaxy
is known or suspected to be a late-type/spiral object, and therefore to have a strong ionized gas component. However,
we are currently interested in the passage of radio waves along a line very close to the centre of a lens galaxy,
and which is therefore potentially subject to a large contribution from propagation effects. With an asymmetric lens
such as 1030+074, the distance from the centre is likely to be at least a few tens of parsec (see e.g. Fig.\ref{geometry}).

In the regime of refractive scattering in which we are interested\footnote{See Rickett (1977) for a full physical
description of different scattering regimes which occur in astrophysics.}, the total scattering effect can be described
by the scattering measure,  SM, using a path integral along the line of sight:

\begin{equation}
{\rm SM} = \int C_{\rm n}^2 ds,
\end{equation}

\noindent where $C_{\rm n}^2$ is the normalization of a turbulence spectrum with a power law spectrum in wavenumber for the
probability of density fluctuations (Cordes \& Lazio 2003),

\begin{equation}
P(k) = C_{\rm n}^2k^{-\alpha},
\end{equation}

\noindent as a function of spatial frequency $k$. The spectral index of this turbulence is usually assumed to follow a 
Kolmogorov spectrum, for which $k=-11/3$.

The quantity $C_{\rm n}^2$ depends on the physical properties of the medium, such as the outer scale of the Kolmogorov turbulence, and the
filling factor and fractional variance of the turbulent clouds. However, it can be shown (Taylor \& Cordes 1993;
Walker 1998; Cordes \& Lazio 2003) that the size of a refractively scattered image is given by

\begin{equation}
\theta = (128~{\rm mas}) {\rm SM}^{3/5} \nu^{-11/5},
\end{equation}

\noindent where $\nu$ is the observing frequency in GHz and SM the scattering measure in kpc$\,$m$^{-20/3}$ (Cordes \& Lazio 2003). 
At the 22-GHz frequency of our VLA observations, this implies that $\theta = 0.14~$mas~SM$^{-3/5}$. In order to scatter
a potential third image to the point where we do not see it, we require $\theta>100$mas, and hence a lower limit of
55000~kpc$\,$m$^{-20/3}$ for the scattering measure.

Detailed models of electron content in elliptical galaxies do not currently exist. In our own Galaxy, we have information
mainly from observations of pulsars throughout the Galaxy, and towards the Galactic centre from pulsars and OH maser
observations. van Langevelde et al. (1992) study the broadening of maser spots within a few tens of parsecs of the
Galactic centre, finding typical extension of a few hundred mas at an observing frequency of 1.6~GHz. This implies 
SM$\sim 100$, well short of the level required to affect our observations, unless the lensing galaxy in B1030+074 has
a radically higher turbulent electron content. Moreover, we note that Winn et al. (2003) present VLBA observations of
the central component of PMNJ1632$-$0033 at a range of frequencies, none of which appear to show the effects of 
refractive scattering. Among these images is a map of the region of the central image with a restoring beam of
10~mas$\times$6~mas, observed
at 1.6GHz; assuming that it really is an image of the background object, its presence requires $SM\leq 1$~kpc$\,$m$^{-20/3}$.

The second possible propagation effect in the lensing galaxy is free-free absorption, which is discussed by Winn et al. (2003)
and Mittal et al. (2007) for the cases of PMNJ1632$-$0033 and CLASS~B0218+357 respectively. Like refractive scattering,
this effect is also frequency dependent, with an optical depth $\tau_\nu$ given by

\begin{equation}
\tau_{\nu}=\left(\frac{\nu}{\nu_{\rm c}}\right)^{-2.1},
\end{equation}

\noindent where $\nu_{\rm c}$ includes the dependence of the optical depth on electron temperature, which goes as $T_{\rm e}^{-1.35}$, 
and a linear dependence on emission measure. Here $\nu$ is the frequency at the redshift of the lens galaxy, which is the
observed frequency multiplied by a factor $(1+z_{\rm l})$.

In the case of PMNJ1632$-$0033, Winn et al. find a best fit of $\nu_c$=3.2~GHz. In the rest frame of the B1030+074 galaxy,
our VLA observation is at a frequency of 35~GHz. Again if we assume that the image detected by Winn et al. is a third
image of the lensed object, we would not expect a significant effect of free-free absorption on our observations, unless
the emission measure were a factor of 10 greater or the electron temperature were implausibly high. Specifically, we
would require an emission measure of $\geq 2\times 10^9$cm$^{-6}$pc to have a significant impact on flux density at 
35~GHz. This is a factor of about 100 greater than that expected from passage through a single H{\sc ii} region. It is also
greater than that expected for passage through a quasar narrow line region; here the densities of the emitting regions
are typically a few hundred cm$^{-3}$, so an implausibly large path length would be needed to obtain the required
emission measure. Therefore, we do not expect propagation effects to have changed the properties of the central image.

\end{document}